\documentclass[journal]{IEEEtran}

\usepackage{subcaption}
\usepackage[pdftex]{graphicx}
\usepackage{adjustbox}
\usepackage{xcolor}

\usepackage{titlesec}
\titlespacing{\section}{0pt}{12pt plus 4pt minus 2pt}{0pt plus 2pt minus 2pt}
\titlespacing{\subsection}{0pt}{12pt plus 4pt minus 2pt}{0pt plus 2pt minus 2pt}
\ifCLASSINFOpdf
\else
\fi
\begin{document}
\title{Realizing Molecular Machine Learning through Communications for Biological AI: Future Directions and Challenges}

\author{Sasitharan~Balasubramaniam,~\IEEEmembership{Senior Member,~IEEE,}
        Samitha~Somathilaka,~\IEEEmembership{Student,~IEEE,}
        
        Sehee~Sun,
        Adrian~Ratwatte,~\IEEEmembership{Student,~IEEE,}
        Massimiliano~Pierobon,~\IEEEmembership{Member,~IEEE}
        % <-this % stops a space
\thanks{{S. Balasubramaniam, S. Somathilaka, S. Sun, A. Ratwatte, M. Pierobon are with the School of Computing, University of Nebraska-Lincoln, NE, USA email: (sasi, ssomathilaka2, ssun12, aratwatte2, maxp)@unl.edu} .}% <-this % stops a space
\thanks{S. Somathilka is also with Walton Institute, South East Technological University, Ireland.}% <-this % stops a space
}

\maketitle

\begin{abstract}
Artificial Intelligence (AI) and Machine Learning (ML) are weaving their way into the fabric of society, where they are playing a crucial role in numerous facets of our lives. As we witness the increased deployment of AI and ML in various types of devices, we benefit from their use into energy-efficient algorithms for low powered devices. In this paper, we investigate a scale and medium that is far smaller than conventional devices as we move towards molecular systems that can be utilized to perform machine learning functions, i.e., Molecular Machine Learning (MML). Fundamental to the operation of MML is the transport, processing, and interpretation of information propagated by molecules through chemical reactions. We begin by reviewing the current approaches that have been developed for MML, before we move towards potential new directions that rely on gene regulatory networks inside biological organisms as well as their population interactions to create neural networks. We then investigate mechanisms for training machine learning structures in biological cells based on calcium signaling and demonstrate their application to build an Analog to Digital Converter (ADC). Lastly, we look at potential future directions as well as challenges that this area could solve.
\end{abstract}


% \fi
\IEEEpeerreviewmaketitle

\section{Introduction}
\IEEEPARstart{I}{n} recent years we have started to witness the widespread development of systems to apply Artificial Intelligence (AI) and Machine Learning (ML) to very diverse application scenarios \cite{jain1996artificial}. This has resulted in software-based systems for AI, such as Artificial Neural Networks (ANN) \cite{mcculloch1943logical} as well as hardware based systems like neuromorphic hardware \cite{bohnstingl2019neuromorphic}. In particular, within the area of ANN various algorithms have been developed, that includes Recurrent Neural Networks (RNN), Convolutional Neural Networks (CNN), amongst others, where each has its own properties and behaviour derived from specific functions of neuronal networks of the brain.  While developments have been made in AI for both hardware and software, there is still a number of challenges that exists. These challenges include the ability to mimic the behaviour and realism of neurons and their internal functionalities, as well as matching their energy requirements. The former challenge is still today a major issue that continues to motivate research to ensure that new algorithms or hardware designs will resemble the properties of internal neuronal signaling (e.g., ion transfer, action potential generation and propagation). However, the more realistic we design AI algorithms to closely resemble neuronal cells, the higher the energy consumption since we are mimicking the chemical and molecular reactions that occurs internally. 
%This in turn leads to the second challenge, which is the ability to ensure that AI technology matches to the energy profile and efficiency of an organism's brain. 
When making this comparison, the brain consumes approximately $20W$ for 100 billion neurons and 1,000 trillion synapses compared to a neuromorphic processor such as the  Neurogrid with 65 thousand neurons and 500M synapses, which consumes $3.1W$ \cite{liu2021low}. In order to minimize energy consumptions, alternative materials have also been proposed for artificial neural systems and one example is the use of spintronics \cite{hirohata2020review}. 

A number of alternative solutions have also been proposed to mimic natural neuron functions, where biological neuronal cells have been used to perform AI computing to  replace conventional computing systems, i.e., biological AI. Examples of this include living neurons that can play pong \cite{kagan2022vitro}, robots integrated with neuronal cells to control their operation \cite{warwick2011experiments}, as well as the control of a robotic arm \cite{bakkum2007embodying}. This approach has also shown that the neurons can also be taught and trained to adapt to specific applications. Besides neurons, other forms for biological systems have also been considered to perform computing functions. Examples include the use of \emph{Physarum} to solve networking problems at the Tokyo railway network \cite{tero2010rules}, and most recently the use of fungii to perform molecular computing \cite{roberts2021mining}. %In each of these works, a fundamental driver to enable learning as well as computing is communication of molecules. 
Using these approaches can possibly result in new solutions where biological cells work in tandem with silicon technologies, i.e., bio-hybrid AI. While this may address the aforementioned challenges of including more realistic biological properties, protocols and technologies to maintain biological cell lines and keeping them alive for a long period may also invalidate the quest for higher efficiency of these systems. %Recent years has started to witness new emerging directions, where other cell lines are used to produce perceptrons like neural network. One example is the use of bacteria to perform AI like computing or organism that dont have a brain or neural system. 

\begin{figure*}%
\centering

\includegraphics[width=1\textwidth,  trim = {0 0 0 0}, clip]{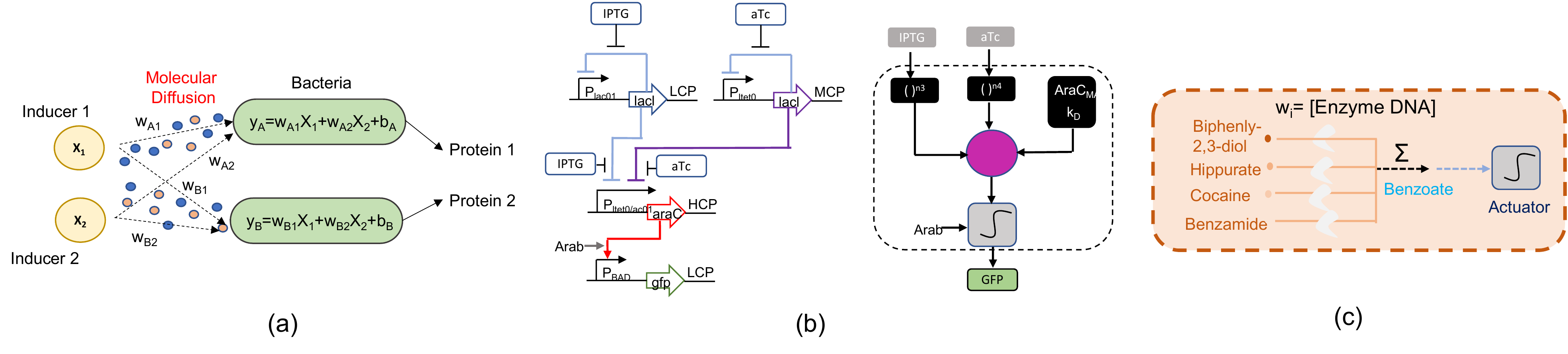}%

\caption{Proposed solutions to develop neural networks from engineering cells, (a) Bactoneuron \cite{sarkar2021single}, (b) Perceptgene \cite{rizik2022synthetic}, and (c) metabolic perceptron \cite{pandi2019metabolic}.}
\label{fig:proposed_approach}
\end{figure*}

Fundamental to all biological AI solutions and models that have been proposed is the exchange of molecules between cells to realize computing functions. This communication based on molecules occurs as both an intra as well as inter-cellular signaling. However, the training and computing processes within these systems can be further enhanced through modeling, optimization, and engineering of these same processes, with the help of molecular communication theory. As this field is slowly maturing, models and systems have been developed to study and engineer  information encoding into molecules to be exchanged between different biological or bio-hybrid entities, also called bio-nanomachines, such as the aforementioned AI-enabling cells. Examples include characterizations of channels within biological environments \cite{pierobon2010physical}\cite{jamali2016channel} \cite{guo2016molecular} \cite{wu2020signal} and molecular modulation techniques (e.g., MoSK \cite{chen2020generalized}). These new communication models have been applied to characterize and engineer numerous types of molecular communication systems such as neuronal interconnections  \cite{bicen2016linear},  multi-hop diffusion-based networks \cite{ahmadzadeh2015analysis}, and large scale systems with 3D geometry \cite{deng2017analyzing}. Test beds and proofs-of-concept have also been developed including table top molecular communication systems \cite{farsad2017novel}, as well as molecular modulators that transmit digital information between computers \cite{grebenstein2018biological}. The engineering of molecular communication systems in biological or bio-hybrid AI systems can enable new design as well as efficiency and robustness. This may include the design of engineered molecules to propagate information during gene expression leading to intra-cellular signaling, as well as inter-cellular signaling that can support ANN functionalities between populations of cells. This can be achieve through the combination of molecular communication theory and the tools provided by synthetic biology, where genetic circuits are engineered to produce molecular signals communicated between cells. 

In this paper, we will analyze a number of different biological AI and the types of communication that is inherent in the models, i.e., Molecular Machine Learning (MML). MML in here intended as machine learning realized with molecules and chemical reactions as building blocks, rather than computer programs to inform synthetic chemistry, as in \cite{pfluger2020molecular}. This includes engineered cells to create perceptrons found in ANN or interconnecting engineered cells to behave as neural networks. We will then follow with alternative future directions for developing ANN using the concepts of molecular communication theory through the natural Gene Regulatory Networks (GRN), molecular communication between multi-species population of cells, as well as engineering of $Ca^{2+}$ signaling based molecular communications to create an Analog-to-Digital Converter (ADC). Lastly we will focus on future challenges for MML.  

This paper is organized as follows. Section \ref{CellAnn} discusses current background on engineered cells as well as metabolic reaction models to realize ANN. In Section \ref{grai} we propose a new direction whereby natural GRNs and their embedded intracellular molecular communication for AI. In Section \ref{multi-bacNN} we introduce an idea for utilizing a multi-species cellular consortia to perform AI using inter-cellular molecular communication. In Section \ref{Capercep} we move towards engineering calcium ($Ca^{2+}$) signaling in cells to achieve perceptron like behaviour. In Section \ref{future} we discuss future directions and challenges, while in Section \ref{conc} we conclude the paper.

%\section{Computing Paradigms in Synthetic Biology}

%The field of synthetic biology aims to develop digital as well as analog computing systems within living cells \cite{cameron2014brief}. As examples, this has led to development of simple logic gates {\bf [REF]} that has been integrated into complex circuits {\bf [REF]}, as well as storage systems {\bf [REF]}. In all these implementations, a fundamental process is communication of molecules to realize these computing functions. The molecular communications can range from short-range intra-cellular signals from expressions of genes to medium-range communication between population of cells. Recent years has witnessed progress towards the impact of natural learning that occurs in biological cells, and how this can be implemented in synthetic computing paradigms of living cells. A number of factors contributes to this. Firstly, biological cells have their own inherent properties that enables them to adapt to the varying environmental conditions. This is a hidden property that is operating in the background, while synthetic engineering of cells are performed to develop computing systems. Secondly, the result of these properties can result in fluctuations and randomness due to the nonlinearity in cellular functions, such as chemical reactions in gene expressions. Therefore, by understanding and linking to these properties, we can (1) develop more stable and robust computing systems if we can integrate natural learning process, and (2) develop novel living machine learning systems, which is the focus on this review.

\section{Current Background on Biological AI}
\label{CellAnn}

Numerous research have indicated natural intelligence that occurs within cells. From the perspective of molecular communications, this deals with initially sensing molecular signals from the environment, followed by  internal signal transduction that leads to gene expressions, as well as corresponding metabolic pathways. This process is largely programmed into the cell's genome \cite{liberman1996cell}. In certain cases, this intelligence and memory management can be performed with organisms that lack a brain, or non-neuronal systems as pointed out in \cite{yang2020encoding}. In the case of bacteria, claims have been made the microbes contain 'minimal cognition' \cite{becerra2022computing}. 

In \cite{sarkar2021single} a single layer ANN was developed using engineered \emph{E.Coli}, known as {\bf Bactoneuron} (Figure~\ref{fig:proposed_approach} (a)). The developed model is able to achieve both reversible as well as irreversible computing. Each cell is engineered to receive inter-cellular diffusing molecules, and as a response, execute a log-sigmoid activation function to produce Green Fluorescent Protein (GFP) output. This execution is established through a transcriptional regulation which is undertaken by an engineered genetic circuit (also referred to as \emph{cellular device}). The solution proposed uses established set of general rules to map the complete ANN architecture and to derive unit bactoneurons directly from the functional truth table of a complex computing function. %developed straightforward anduniversal molecular design rules directly from the mathematicalnature of the activation function, which connect themolecular design of an individual cellular device and the sign ofweights in an activation function of a bactoneuron in a one-to-onefashion. 
The study produced both simulations as well as experimental validation. Example applications included a 2-to-4 decoder, a 4-to-2-priority encoder, a majority function, a 1-to-2 de-multiplexer, and a 2-to-1
multiplexer and reversible logic mapping through Feynman
and Fredkin gates.% {\bf**check weights for this work}.

Rizik et al \cite{rizik2022synthetic}  developed the {\bf Perceptgene} (Figure~\ref{fig:proposed_approach} (b)), which is a perceptron model of an ANN. This was achieved through the genetic circuit engineering in \emph{E. Coli} bacteria. The perceptron behaviour is established through a logarithmic input-output relationship that fits to the non-linear biochemical reactions that occur in the genetic circuits. The implementation is based on engineered genetic circuits whose input-output behavior includes both the power-law as well as a multiplication function. The power-law function encodes the weighted chemical inputs, while the multiplication function aggregates all the inputs that will determine the activation. The weight of each input is determined by the Hill coefficient. The two inputs used are \emph{isopropyl Beta-D-1-thiogalactopyranoside} (\emph{IPTG}) and \emph{anhydrotetracycline} (\emph{aTc}) molecular signals and results in a repression process that in turn regulates their own production using an auto-negative feedback loop. Similar to the perceptrons of an ANN, the perceptgene also contains a bias component for the sigmoidal activation function. The bias input is set by the ratio of the maximum transcription process to the binding affinities of the protein-protein/protein-DNA reactions. The applications of the perceptgene include weighted multi-input functions, classification, as well as an offline gradient descent  learning algorithms. %\textcolor{blue}{Comment: we have to say that this is offline as well, I think...} The weighted inputs are determined by the Hill coefficient, where the linear combination of the inputs conducted in the logarithmic domain. A central property that perceptgene operates on is the non-linear biological pathways that occurs in gene regulations, which exhibits logarithmic and power law input-output relations (here the outcome is dictated by the level of fold-changes rather than absolute level of outputs). Using this, the perceptgene is able to perform logarithmic input-output operation that suits the nature of gene regulation process. 

In \cite{becerra2022computing}, an offline trained perceptron neural network is used to program a population of bacteria and it is simulated \emph{in silico}. Through the diffusion of inter-cellular molecular communication within a population, the cells were able to have social interactions and form complex communities. The programmed perceptron was also used to solve an optimization problem. The work was based on an in-silico model, where the plasmid encoded perceptron was designed using \emph{Cello}, while the simulation of the bacterial communication was developed through the \emph{Gro} simulation tool. A particular aspect of the study is the use of programmed ANN into the genetic circuit to control signaling between cells in the population to perform  functions. The input are natural molecules (e.g., galactose), which in turn control a downstream behaviour. This includes (i) emitting molecular signals proportional to the concentration of oxygen that is used for metabolic purposes, (ii) inducing chemotaxis for cell movement, (iii) commensalism, where the cells emit a signal that degrades the waste products from other bacteria in the population, and (iv) controlling of cell growth when the environment is harsh.

In \cite{li2021synthetic}, a consortia-based bacterial ANN was developed and proved experimentally. An interesting feedback process is developed between the receiver and the sender, which are the perceptron nodes for decision making and this is achieved using quorum sensing. The sender bacteria are able to emit varying molecular signals (\emph{OHC14} - \emph{acyl-homoserine lactone 3OHC14:1-HSL}), which represent the weights. These molecular signals are induced by an external signal (\emph{OC6} (\emph{acyl-homoserine lactone 3OC6-HSL})). The application was specific to 4-bit pattern recognition, where varying levels of the \emph{OC6} inducers are applied to sender bacterial populations and once the molecular signals diffuse to the receiver, they will activate a genetic circuit to produce an output signal. A novel gradient descent algorithm was also developed to optimize the weights of molecular signals to suit the pattern recognition application.  

A cell-free perceptron model was proposed in \cite{pandi2019metabolic} using the metabolic circuit illustrated in Figure~\ref{fig:proposed_approach} (c). The latter was designed with a focus on biochemical retrosynthesis to predict the pathways, which was achieved using the \emph{Retropath} and \emph{Sensipath} computational design tools. The circuit was then embedded into a cell-free system in order to create the {\bf Metabolic Perceptron}. The metabolic perceptron was able to perform binary classification based on metabolite molecular signals that leads to a classification process. The example application was here a four-input binary classifier. 

\section{Genetic Regulatory AI}
\label{grai}
While the previous section focused on the genetic engineering of living cells to create machine learning systems, in this section we will look at an alternative approach that is based on computing structures naturally present in biological cells, i.e., GRNs. This approach is based on essential similarities between a GRN and its structure to an ANN. %This is a difference compared to previous works that has focused on characterizing GRN building on the non-linear relationships that occurs in biochemical reactions, as well as we build on this by determining the sequence of gene expressions that results in a logarithmic behaviour. {\bf** do we see logarithmic and power law input-output relationships when there is sequence of gene expression?} 
While a number of different works have investigated neural-like properties in GRNs, our investigation focuses on how molecular communication properties can be exploited to perform computing functions as well as training by externally manipulating the weight connections between gene relationships. %In this form of learning process, the analog input is based on the non-linear process of gene expressions ({\bf **this will be collective non-linear gene expressions - what does this result in?}) that triggers an activation function, which can be sigmoidal or step-function. Essentially, this will result in integration of molecular signals {\bf**how does multiple molecular pulses that disperses and spreads gets integrated at each node serves as input to the activation function,  and what is this relationship to activating molecular signals}).  

%creating sum temporary subsections for easier organizing

\subsection{Background on Gene Regulatory Networks}
A GRN is a highly complex network of multi-layered interactions between genes. Each individual cell carries a GRN specific to its species and strain, giving an unique behavioural pattern as well as functionalities. A cell can sense a range of external stimuli using membrane receptors, perform computing through the GRN and express genes accordingly, thus resembling an input-process-output sequence found in conventional computing. A typical process of gene expression starts with the transcription process of converting the genes into mRNA and this depending on the gene can be followed by the translation process that coverts the information contained in the mRNA into proteins. However, during gene expression within the GRN, molecular communication patterns can be identified in gene-gene interactions, which are complex processes that occur at multiple layers. For example, while these interactions in prokaryotes contribute to the regulation of the aforementioned transcription process, for eukaryotes they can be post-transcriptional, i.e., contributing to, among other things, mRNA (or other transcript) and/or protein functionalities.

Moreover, the regulation in the post-transcription layer contributes to specific dynamics in the behavior of GNRs. In this context, proteins plays a crucial role complementing the regulation mechanism by integrating sensing, transfer, storage, and processing of information. 
%In terms of sensing, proteins are capable of receiving a range of external stimuli such as light, temperature, mechanical forces, or voltage and respond by the formation of a macromolecular structure, the generation of a physical movement, or the production of light. The information transfer from receptors to the actuators is extremely rapid, where in some instances, such as receptor to flagellar motor takes less than 0.1s, and protein-built voltage-gated or ligand-gated ion channels can change from one state to another in less than 50$\mu$s.
As an example, proteins can perform computational tasks such as amplification, Boolean logic functions, and information storage through mechanisms of allosteric regulation \cite{bray1995protein}.
%In some cases, the logic gate behavior of proteins can also be changed based on inputs, thus advancing toward fuzzy logic. 
In addition, the inter-conversions between phosphorylated and non-phosphorylated states of proteins act as switches enabling them to exhibit sigmoidal behaviours over a limited concentration range. 
%Therefore, factors ranging from the concentraion of TFs, to the allosteric regulation of protein influence the impact from one geneto another. There are representatives model of GRN using Boolean logic circuits that can be constructed from kinetic equations and binding equilibria, Bayesian networks, fuzzy logic, and Neural networks \cite{vohradsky2001neural}. The models usually integrate the probabilistic behaviour of the regulation mechanism, which reflect the uncertainties entangled in the gene expression, such as inherited stochasticity of gene expression and intracellular-MC dynamics of transcriptional factors \cite{shmulevich2002boolean}. 

In the following, we show how these complex molecular signaling processes that involve multiple layers of chemical reactions as well as components during gene expressions, combined with the network structure of genome relationships, can allow us to identify and exploit natural ANN within GRNs, i.e., Genetic Regulatory AI (GRAI).

\begin{figure*}
\centering
\begin{tabular}{cc}
\adjustbox{valign=b}{\subfloat[\label{fig:Intro_GRN}]{%
      \includegraphics[width=.6\linewidth]{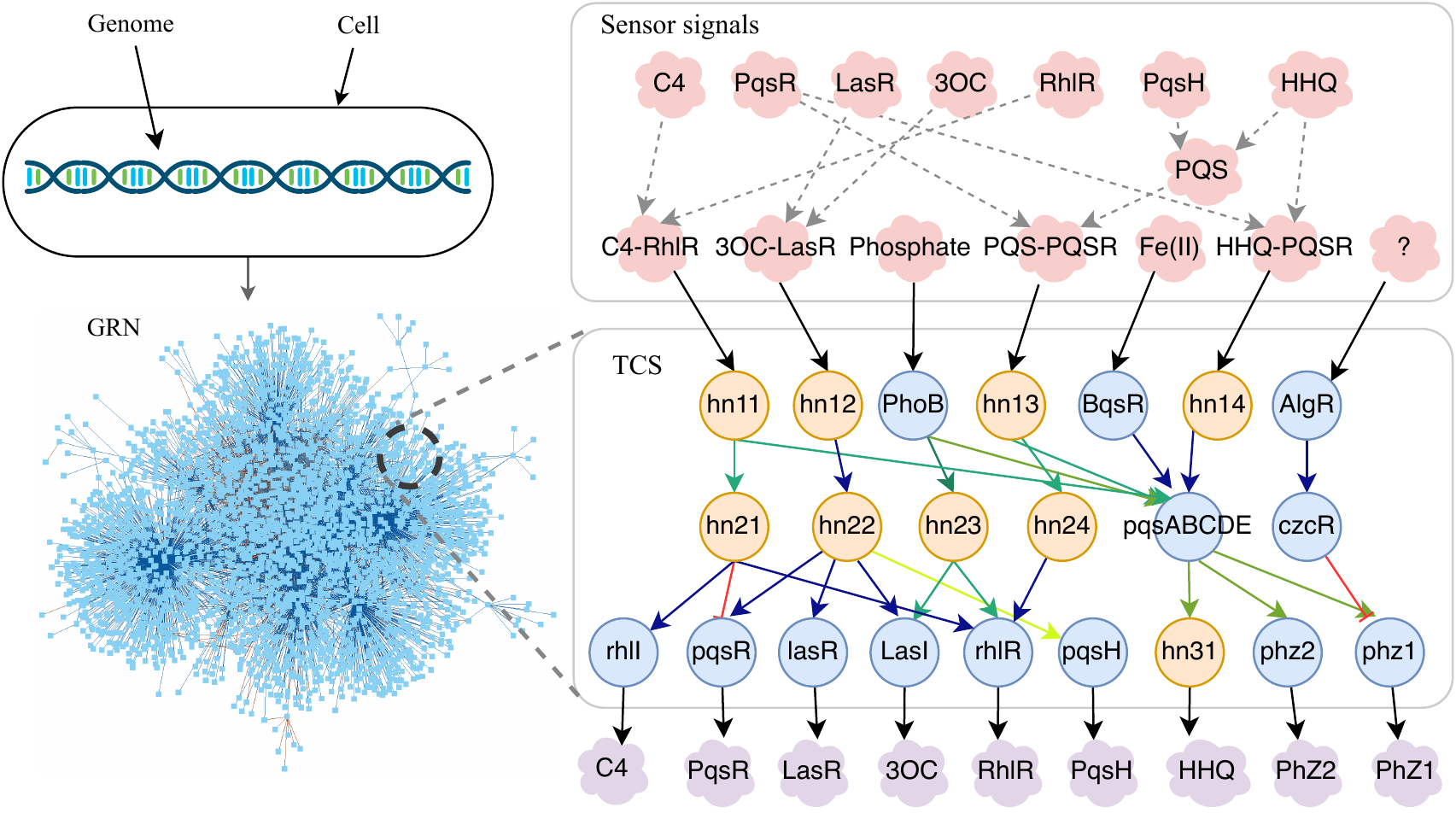}}}
&      
\adjustbox{valign=b}{\begin{tabular}{@{}c@{}}
\subfloat[\label{fig:37NN}]{%
      \includegraphics[trim={0 145 0 0},clip,width=0.38 \textwidth]{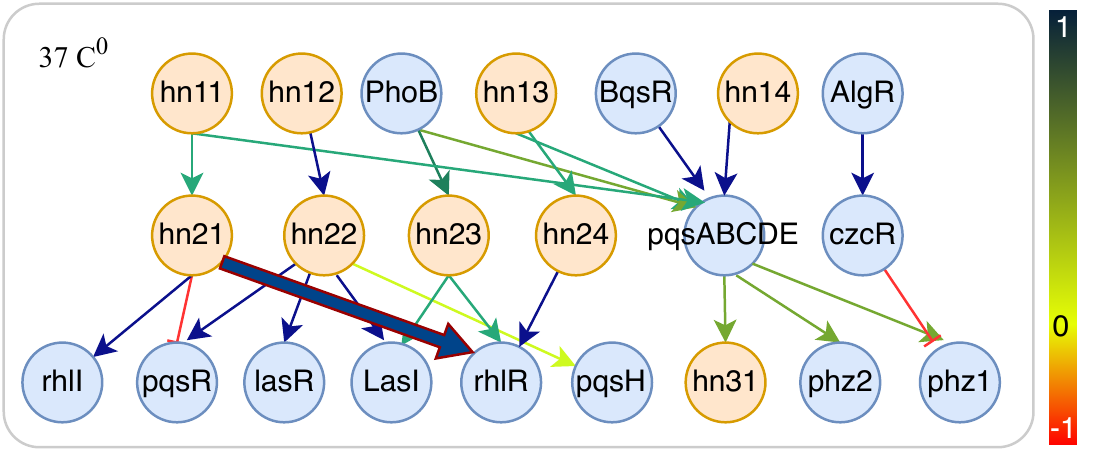}} \\
\subfloat[\label{fig:30NN}]{%
      \includegraphics[trim={0 0 0 140},clip,width=0.38 \textwidth]{Diagrams/Intro_b.pdf}}
\end{tabular}}
\end{tabular}
\caption{Illustration of inherited GRAI where, a) shows the extraction of a subnetwork that resembles an ANN with relative weights, b) set of relative weights in one environment condition (temperature at 37$^0C$), and c) modified weight in a different environment condition (temperature at 30$^0C$). }\label{fig:TCS_GRAI_extraction}
\end{figure*}

\subsection{ANN Learning and Training Models in a Simple Gene Regulatory Network}

The transcription of a particular gene in a GRN is combinatorial action of products of other genes as well as its own. Subsequently, the state of the cell is an action based on a combination of diverse translated gene products. When we observe these properties, we see a resemblance to the dynamics of an ANN,  specifically a Recurrent Neural Network (RNN), where the current state depends on the previous. This means that there is a potential to create MML from manipulating the gene expression patterns.% based on their established relationships.
%Nevertheless, there were different approaches to capturing the dynamics of GRNs, such as 

%\section{Bacterial Systems}

%A typical bacterial cell is sensitive to a wide range of externals signals such as temperature, pH levels, presence of other microbial organisms and a large number of metabolites. These external stimuli as signals, trigger gene expressions of the cell forming mesh of intergenic MC signals in the GRN. Hence, we explored the idea of GRAI in bacterial cells focusing on a subset of genes that are associated with TCSs in this section.
To describe our concept, we will focus on a simple communication pattern found in the GRN of a bacterial cell. Bacteria uses signal transduction pathways to sense the environment by processing input signals. \emph{Two-Component Systems} (\emph{TCS}) are among the most widespread signal transduction mechanisms, which contain a \emph{Sensor Histidine Kinase} (\emph{SHK}) that receives external signals and a response regulator that accordingly initiates the expression of a set of genes. On average, a bacterial cell contains 30 TCSs that are essential for their virulence, growth and survival. Approximately, 87\% of the known response regulators of TCS involve gene expression regulation at the transcription layer. Based on this, 96\% of SHKs are capable of sensing small-molecule-binding from the extra-cellular space. Hence, the combination of TCSs can be considered a viable example of a natural GRN pattern that can be modeled and characterized as an ANN, where the input layer is represented by the SHKs, and multiple hidden layers as well as an output layer consist of genes and their mutual interactions. There are several advantages in using the TCS sub-network of the GRN as an ANN for MML. This includes availability of experimental data that offer validation and quantification of the relationships between gene expressions for both input and output layers. In a number of cases, the direct mapping of a GRN sub-network to an ANN is not feasible. The reason is because sometimes the number of gene interactions (network hops) from the input layer to the output layer can vary for different gene expression paths, resulting in the corresponding ANN to be asymmetric, which leads to less computational efficiency. There are well-known approaches to address this problem, such as introducing phantom nodes that do not alter the overall behavior or treat the network as asymmetric ANN structure. Another alternative is to introduce missing gene interactions through engineered genetic circuits, which can further align the sub-network closer to a typical ANN structure.  
% \begin{figure*}
% \centering
% \begin{subfigure}{.97\columnwidth}
% \includegraphics[width=1.1\columnwidth]{Diagram_Adrian/Diagram_IEEE_rev.pdf}%
% \caption{}
% \label{fig:diagram_Adrian}
% \end{subfigure}\hfill%
% \begin{subfigure}{.99\columnwidth}
% \includegraphics[width=\columnwidth]{Diagram_Adrian/Diff_Structures_IEEE(2).pdf}%
% \caption{}
% \label{fig:Subnets}
% \end{subfigure}%
% \caption{Two sub-networks extracted from the GRN is shown in (a) and number of sub-network structures in the GRN with different number of inputs and outputs is illustrated in (b)}
% \label{fig:hasil}
% \end{figure*}

\begin{figure}
     \centering
     \begin{subfigure}[b]{0.5\textwidth}
         \centering
         \includegraphics[width=\textwidth]{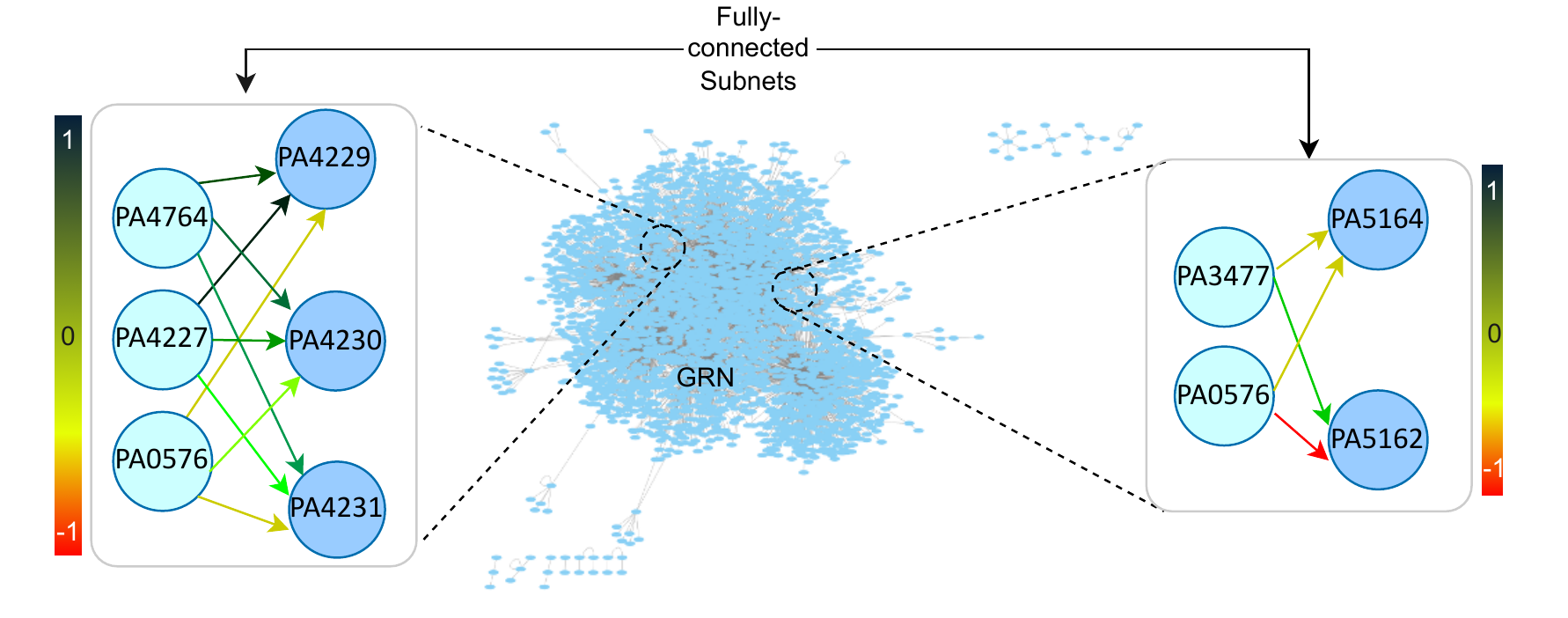}
         \caption{  }
         \label{fig:diagram_Adrian}
     \end{subfigure}
     \hfill
     \begin{subfigure}[b]{0.5\textwidth}
         \centering
         \includegraphics[width=\textwidth, height=6.5cm]{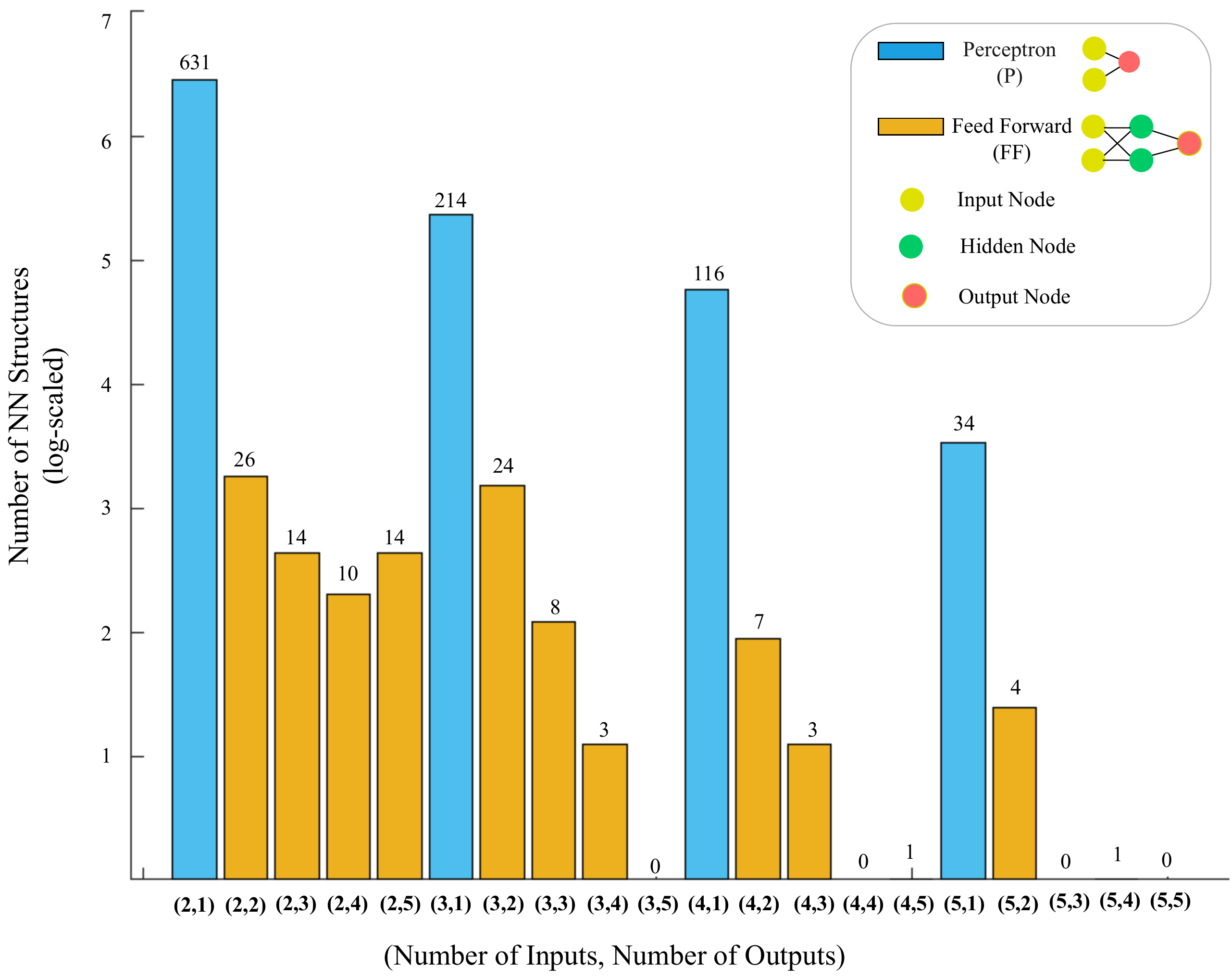}
         \caption{}
         \label{fig:Subnets}
     \end{subfigure}
     \caption{Two fully-connected ANN sub-networks extracted from the full GRN is shown in (a) and number of different sub-network structures that can extracted from the GRN is illustrated in (b)}
\end{figure}

Figure \ref{fig:TCS_GRAI_extraction} illustrates how we recognize an ANN structure from a TCS sub-network of a GRN. As shown in the figure, the  cell is able to combine multiple input signals and accordingly express downstream genes through the network. Gene expression products from one gene reach the non-coding region of another via intra-cellular diffusion \cite{schavemaker2018important}. The relationship of genes to be expressed in the network can be associated to a set of weights. The values of the weights are a result of several factors that include the transcription factors, affinity of the transcription factor binding site, thermoregulation, enhancers \cite{grosso2014regulation} as well as the noise due to the diffusive motion of regulatory molecules \cite{van2006diffusion, 7405285}. %The modification of the weights in this example is based on the external influence from the environmental conditions. 
Here, we focus mainly on two TCSs: \emph{PhoB-PhoR} and \emph{BqsR-BqsS} systems, which are associated with phosphate and iron uptake of the \textit{P. aeruginosa} species. Further, we target the inter-cellular molecular communications by considering three QS systems, namely, \emph{Las}, \emph{Rhl} and \emph{PQS} genes where \emph{Las} uses \emph{3O-C12-HSL} and \emph{Rhl} uses \emph{C4-HSL}, while the \emph{PQS} relies on \emph{2-heptyl-3-hydroxy-4(1H)-quinolone}. To identify the corresponding ANN structure, we first modeled the GRNs as graphs using the interaction structural data from publicly available database \cite{GalnVsquez2020}. This is followed by extracting the TCS sub-network related to the phosphate intakes iron along with the quorum sensing process. The obtained ANN model contains various numbers of hops from the input layer to the output layer, which require the introduction of phantom nodes that do not have an impact on the interaction dynamics of the network. The weights of the ANN represented by the TCS are estimated relatively using the interaction dynamics as well as transcriptomic data \cite{venturi2006regulation, francis2017two}. %Since the obtained ANN model is extracted from the GRNs are based on gene expressions, this will be associated with natural metabolic network in order to create a realistic model of gene expression. 
The performance accuracy of this model is then evaluated based 
%on the \emph{pyocyanin} production by using several experiments with different nutrient concentrations and mutant species. Mainly, we compared the model's computations on 
the pyocyanin production and gene expression levels in low and high phosphate conditions with the data from wet-lab experiments in similar setups \cite{meng2020molecular}. %Further, we conducted a set of mutagenesis experiments to analyze gene expression behaviors of the extracted NN in-depth. The results of these mutagenesis experiments were also compared with the wet-lab data obtained from previous studies \cite{matilla2022virulence}. These comparisons proved that the NN model computes external signals similar to cells, emphasizing the existence of a GRNAI.

A typical ANN will require modification of weights as its being trained to serve for a specific purpose. Here, we investigated how the weights of the ANN related to the TCS can be changed with a specific focus on changes that can be operated externally to the biological cell, from the environment. Previous research has demonstrated how the temperature can impact the cellular functions of \textit{P. aeruginosa}. This usually results in the modulation of one specific gene expression interaction of the \emph{Rhl} QS system \cite{grosso2014regulation}. As highlighted in Figure \ref{fig:37NN}, with the reception of \emph{C4-RhlR} at 37$^0C$ temperature, the weight of \emph{hn21} - \emph{rhlR} is significantly higher compared to the same at 30$^0C$, as shown in Figure \ref{fig:30NN}. This corresponds to a higher expression rate of \emph{RhlR} at 37$^0C$. This demonstrates that updating and training of GRAIs is possible through changes in the environmental conditions, such as temperature.
%This weight regulation takes place due to the  post-transcriptional process of  repression due to the heat-shock gene expression thermometer {\bf please rewrite this sentence}.

 % {\bf Samitha's work: }

% {\bf Adrian's work:} {\bf**look at different neural network structures and which resembles the most to GRN-AI (Feed-forward neural networks)}

\subsection{Mining ANN in GRNs}
\par

%Artificial Neural Networks (ANN) have been studied broadly and applied in many fields for last two decades. 
Our previous section has shown that certain sub-networks of the GRN exhibit natural neural networks. In this section we want to investigate if other sub-networks that exhibit ANN structures can be extracted from the GRN. We perform this through a search algorithm that mines the GRN for specific types of structures. During the search process, if we need a structure with $i$ number of input nodes and $j$ number of output nodes, the algorithm first mines $j$ number of nodes that have a common predecessor. %For example, PA4764 is a predecessor node of PA4229 illustrated in Figure \ref{fig:diagram_Adrian}. 
The $j$ number of nodes will have a number of different predecessors and will be put together into the same group. Within the same group, the nodes will be put together to create different combination, where the combinations must have $i$ number of input nodes that re the predecessor as well as $j$ output nodes. These combination will reflect the different number of sub-network for nodes input nodes $i$ and output nodes $j$.
 
 %serve thwithout output edges and contains a  given number of input edges {\color{blue}$(i)$}. The selected nodes are then grouped according to their mutual relationship of the predecessor node.  Then, select combinations between them {\bf**what is them? be specific} depending on our number of output nodes.{\color{blue}Finally, we take different combinations between filtered nodes above, within the same group. Which means we consider combinations between nodes that has no output edges, only input edges which is equal to our number of inputs and with equal predecessors. }

\label{popann}

\begin{figure*}
\centering
\includegraphics[width=1\textwidth,  trim = {0 0 0 0}, clip]{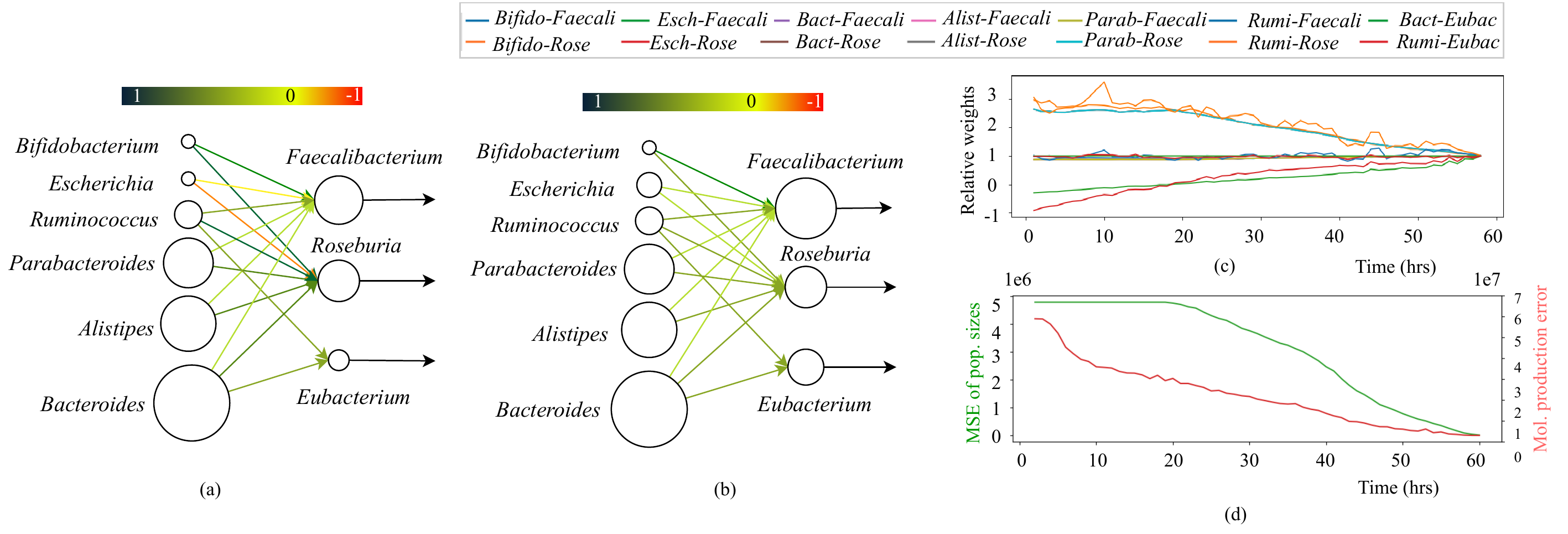}%

\subfloat{\includegraphics[trim={0 0 0 0},clip,width=0.33 \textwidth]{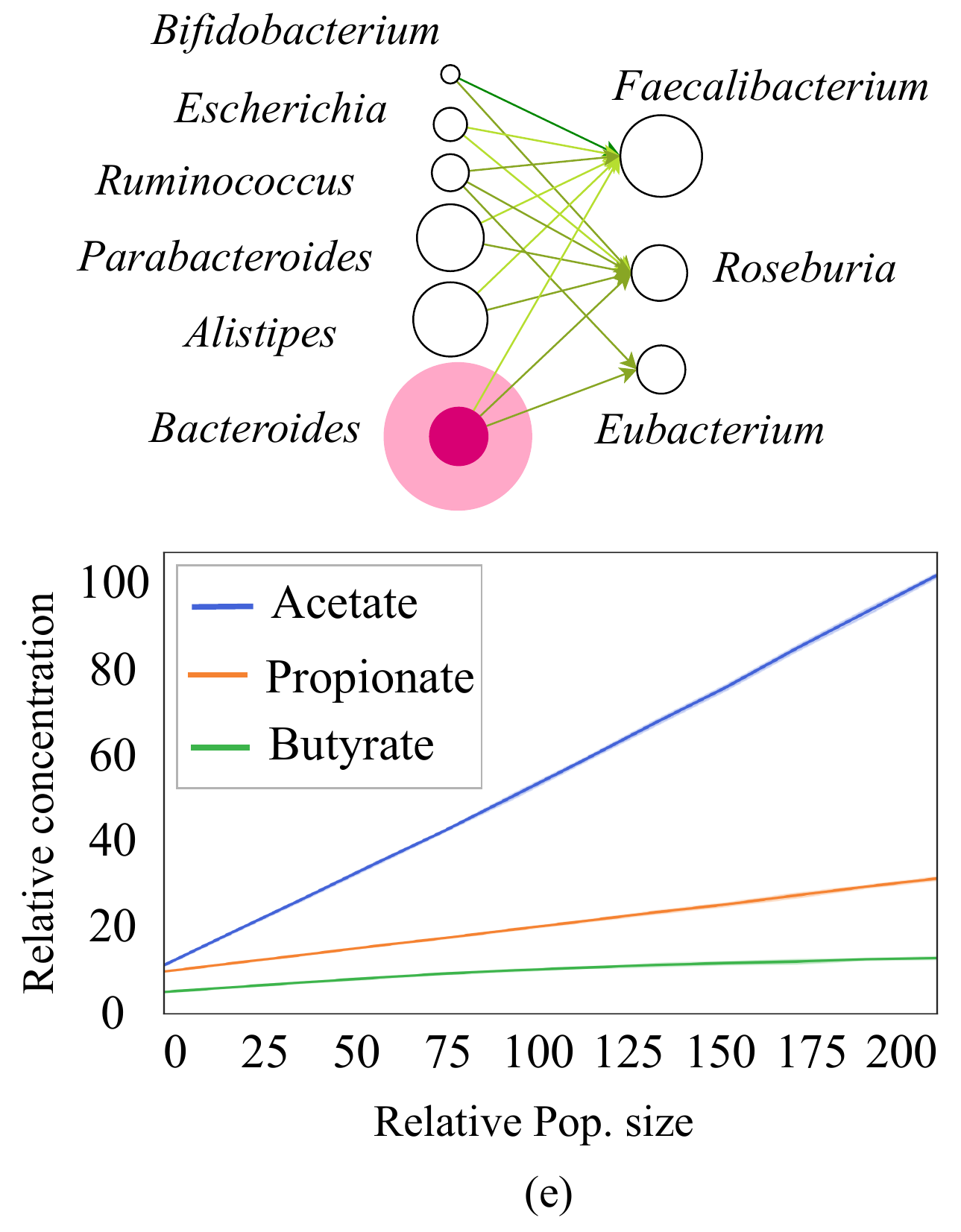}}
\subfloat{\includegraphics[trim={0 0 0 0},clip,width=0.33 \textwidth]{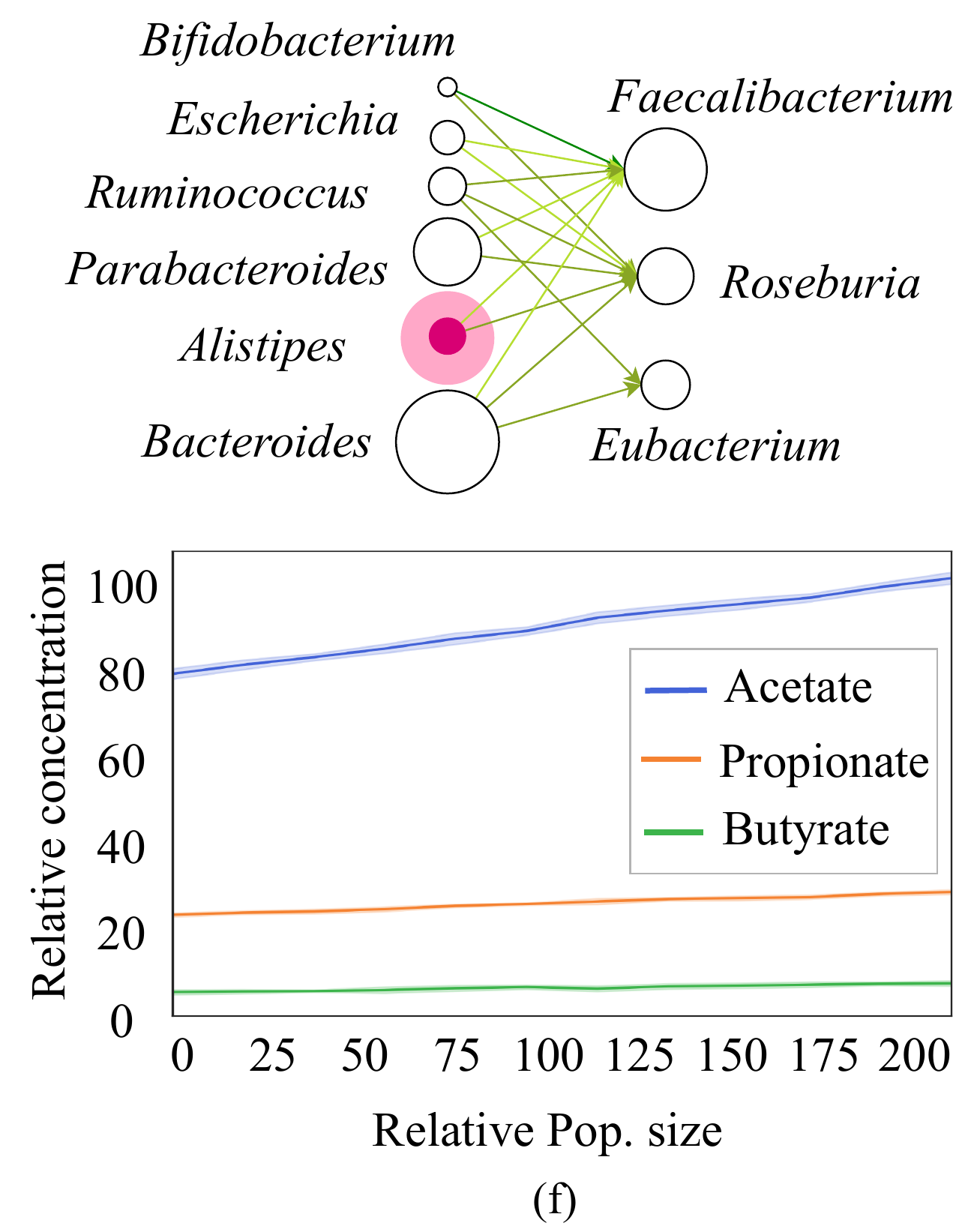}}
\subfloat{\includegraphics[trim={0 0 0 0},clip,width=0.33 \textwidth]{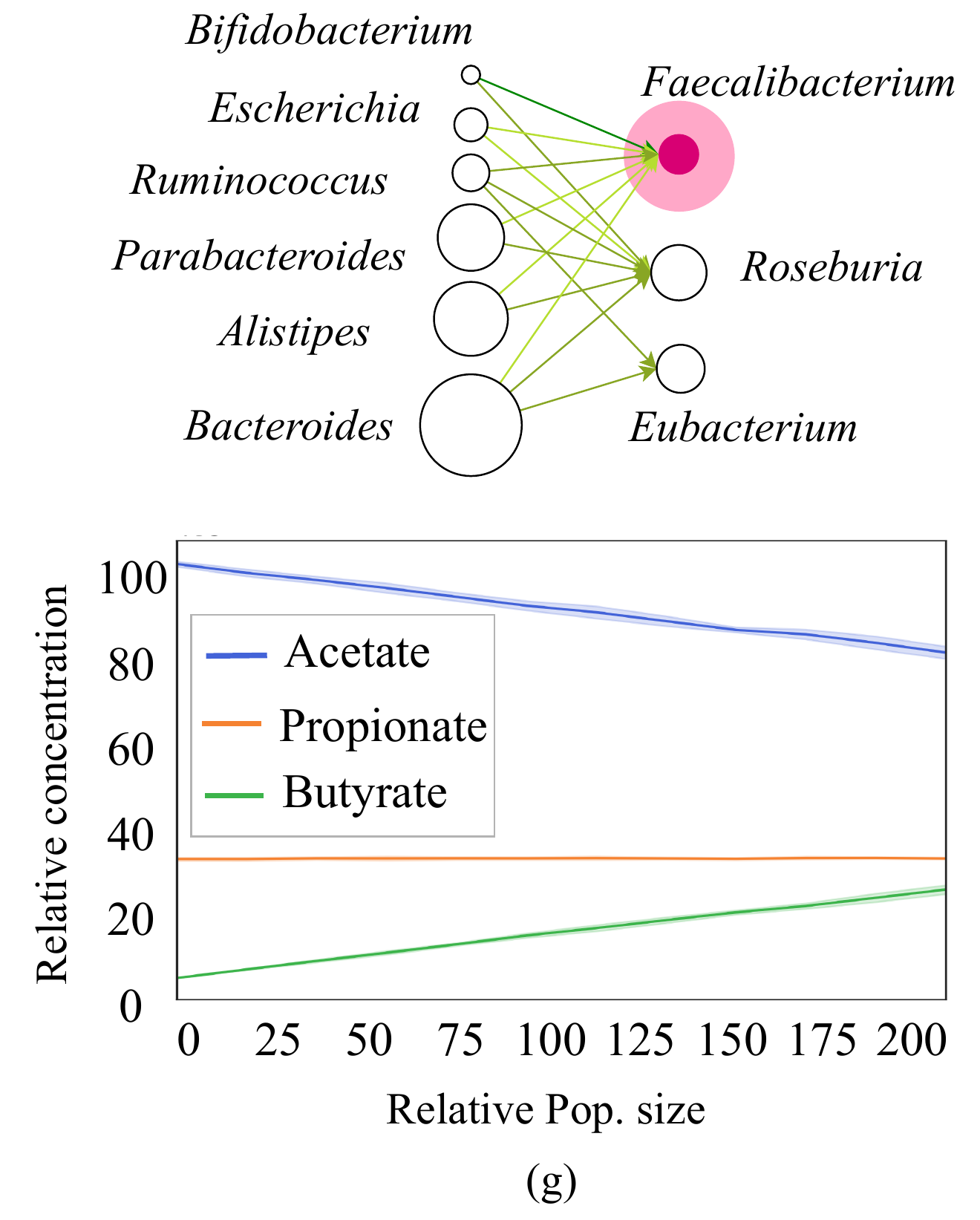}}

\caption{Illustration of population-based ANN weight alteration and its impact on the network outputs is shown here, where a) is the initial ANN setup, b) is the ANN with the preferred network weights, b) is the convergence of weights of all the edges relative to the preferred ANN over the transformation period and d) is the MSE behaviors of molecular production relative to the preferred ANN weights. Further, the output signal behaviors due to variations in weights caused by network structural changes are shown in e), f) and g) by changing the population sizes of \textit{Bacteroides}, \textit{Alistipes}, and \textit{Faecalibacterium}, respectively.}
\label{fig:PopBasedNN}
\end{figure*}
%Since human brain is also consist natural neural networks this study will help us to understand, how human brain processes information.   Our aim is to extract different sub-networks from natural neural networks and to  analyze their AI capabilities. Hence, we will be able to reveal that natural-neural-networks also consists structures that perform AI other than Artificial Neural Networks. 
\par
Figure \ref{fig:diagram_Adrian} illustrates examples of a Feed-Forward neural network with different structures of fully connected ANN sub-networks extracted from the GRN. %As shown, we have considered one structures with two input nodes and two output nodes and another structure of three input nodes with three output nodes. 
Figure \ref{fig:Subnets} shows the number of perceptron and Feed-Forward neural network structures we obtained from the GRN using our mining algorithm. %Perceptron is a structure with two input nodes and one output node while FF can be two-layer or three-layer with a hidden layer. 
We are able to discover a significant number of perceptron structures with the highest  recorded for one output node and two input nodes. As we increase the number of inputs, the number of fully connected Feed-Forward networks becomes harder to discover. In particular, Feed-Forward networks with five output nodes and higher than three input nodes are very rare. 

% Figure \ref{fig:Subnets} shows the number of sub-networks we obtained from the full GRN using our  mining algorithm. For small networks, we are able to discover a number of feedforward NNs. The highest number of structure we found is with one input node and three output nodes. As we increase the number of inputs and outputs, the quantity of fully connected ANN network becomes harder to discover.
% For example, the number of structures with three input nodes dropped below 200 when we increased the number of output nodes. 
% However, there are between 1000 to 25 million structures with only one input and output nodes that are less than 4.
Since these Feed-Forward neural networks are pre-trained with defined weights, the question now rises as to how we can use this for applications. One approach towards using the  ANN found in the GRN is to match it to an application's requirement. This will require a mining algorithm that matches to the problems that require an ANN with the same structure as well as weight combination.
%Based on our results, we can mine for different structures within a GRN and we can use trained weights of the edges to check whether the network acts as ANN. Therefore, we can select a structure depending on weights which matches to our problem. 
While this can create challenges in terms of finding the right problem to suit the ANN found in a GRN, there is an opportunity to engineer the circuit with addition of genes that will increase the diversity of the network as well as integrate hidden layers. 
%As future work, we will mine for different structures with hidden layers and will implement on different GRN. In addition, we will upgrade our algorithm which will improve efficiency while reducing computation time. In addition, we will compare different sub-network structures to find the best fitted network for a given problem.

\section{Bacterial Multi-species Diffusion-Based Neural Network}
\label{multi-bacNN}

In this section, we look at an alternative model for MML, where we investigate how multiple species of bacteria with symbiotic relationships, such as those found in a bacteriome, i.e., bacteria living in endosymbiosis with a host organisms, can be modeled and exploited as an ANN. In general, bacteria of the same species receive specific types of molecular signals from other populations and process them to produce a set of molecules that can influence other species or host cells. These multi-species bacterial populations can be considered the nodes of a network, where the molecular signals that diffuse between population are the link/edges, based on diffusion-based molecular communications. As the molecular signal cascades through the network from layer to layer, this resembles a feed forward neural network (layer in this instance are bacterial species that receive the same type of signals). The relationship structure of the bacteria and signaling weights depend on factors such as the diversity of the species, population sizes, cross-feeding/inter-cellular communications and molecular signal diffusion dynamics. 
The population sizes determine the rate of molecular signal reception and production and this reflects the weight of the edges of the corresponding ANN model. If a larger population produces a signal and another population that has higher relative abundance consumes that signal, the weight corresponding to the link between these larger populations will be modeled with an ANN edge with a larger weight. On the other hand, if the population sizes of the two different species are smaller, the interaction between them is comparatively weaker and will result in a smaller weight value of the corresponding edge.

One of the well-studied bacterial ecosystems is the Human Gut Bacteriome (HGB), which constitutes up to 1000 species \cite{9839035} and it suggests a relevant use case for the aforementioned concept. The reliability of the molecular signal flow between the different species is vital in modeling and exploiting the ecosystem as an ANN. %Hence, using Mutual Information (MI) analysis on the Short Chain Fatty Acid (SCFA) production network, one of our previous studies showed that the HGB is reliable in terms of information flow. Here, considering the metabolism similarities relative to the SCFA production process, we interpreted populations at the phyla level as nodes of the HGB network. 
In our previous study, the structural derivation of a network of multi-bacterial species using graph theory was analyzed, where input of glucose is received by certain species to produce various Short Chain Fatty Acid (SCFA) communicated between the cells \cite{9705067}. The study revealed that the weights of the edges, which are the lactate and acetate signals exchanged between the populations, can be modified and adapted based on external inputs (e.g., glucose). Using this concept, we believe we could design a Bacterial Multi-species ANN from the SCFA molecular communication network within the HGB. Figure \ref{fig:PopBasedNN}(a) illustrates an  example of multi-species bacteria population that are organized into an ANN structure. The arrangement of the structure is based on the input-output relationship of molecular production. For example, when input glucose is consumed, it produces lactate and two SCFA (acetate and proprionate) by six species to produce butyrate for other species, then the six species will be the first layer of a corresponding ANN of our NN, and the species that produce butyrate will be the ANN's second layer. Figure \ref{fig:PopBasedNN}a shows the ANN with the relative weights of each edge shown with different color shades. Our aim is to train the ANN in Figure \ref{fig:PopBasedNN}a into an ANN with a specific functionality, shown in Figure \ref{fig:PopBasedNN}b. Our training is based on the external input of glucose, where we can see in Figure \ref{fig:PopBasedNN}c that as the species are consuming and producing molecules, their weight is slowly being modulated by changing the population sizes, Figure \ref{fig:PopBasedNN}d (as the Mean Squared Error (MSE) of the population converges, similarly the molecular production error). % explain the behaviors of weights and the errors with respect to the preferred NN that is shown in Fig. \ref{fig:PopBasedNN}b. 

Further, we show how significant the impact of the population size variation is on the overall gut metabolic performance by altering the abundance of each species relative to a healthy HGB composition. Figure \ref{fig:PopBasedNN}e shows the network outputs in terms of acetate, propionate and butyrate when the abundance of \textit{Bacteroides} is changed from zero cells in the environment to a population size of 200\% as in the healthy HGB. Figure \ref{fig:PopBasedNN}f and Figure \ref{fig:PopBasedNN}g present the behaviors of the same outputs when altering the population sizes of \textit{Alistipes} and \textit{Faecalibacterium}, respectively. These results indicate the possibility of altering weights of Bacterial Multi-species ANN to modify the network outputs significantly, which can be used in applications such as personalized treatment of metabolic disorders.

%Imbalanced or deviated HGB compositions (dysbiosis) are known to be a threat to human health. An imbalanced SCFA NN may compute the metabolic inputs and produce outputs that can cause various metabolic disorders. Hence, shifting the weights from one state to another preferable state using external influences can be used as a personalized therapeutic approach targeting the unique metabolic needs of the host.

\section{$Ca^{2+}$ Signaling Perceptron Based on Molecular Communications}
\label{Capercep}

In this section we discuss a perceptron that can be trained by controlling the ion flow as well as the basal reactions of $Ca^{2+}$ Signaling between biological cells. As an example, we demonstrate the design of a multi-cell ADC realized by modulating the cell's $Ca^{2+}$ influx as well as through the engineering of genetic circuits. 

\subsection{Calcium Signaling}
Communication through  $Ca^{2+}$ ions is one of the essential signaling processes at the basis of numerous cell functions. %The ions-based signaling often affects the entire cell’s behavior and can play a critical role in its self-regulation as well as death. 
While a few mathematical models for $Ca^{2+}$ signaling have been proposed, the model by Korngren et al. for $Ca^{2+}$ ion transients in electrically non-excitable cells is one of the most recognized and is at the basis of the concepts we present in the following \cite{KorngreenA.1997Armo}. %The model integrates $Ca^{2+}$ ions store within the cytoplasm, the channel within the plasma membrane, membrane receptors, cytoplasmic $Ca^{2+}$ ions buffer, as well as $Ca^{2+}$ ions pumps into the cell. 
According to this well-regarded model, this communication process is based on $Ca^{2+}$ ion influx into the cytoplasm from the extracellular medium, where ion-conducting channels are established through the membrane and controlled by receptors. The receptor in the model is designed in terms of a linear activation instead of complicated non-linear agonist binding curve \cite{KorngreenA.1997Armo}.
%The receptor in the model is designed as a linear activation term rather than an agonist binding curve \cite{KorngreenA.1997Armo}. 
As the influx of ions increases the $Ca^{2+}$ signaling reaction is activated, where the $Ca^{2+}$ ion pumps allow the outflow of ions from the cytoplasm to the external medium as well as its store. %This is followed by the $Ca^{2+}$ ions buffer association with the quantity inside the cytoplasm to control the concentration levels.
Eventually, the $Ca^{2+}$ ions concentration in the cytoplasm reaches a saturated level. Based on this sequence of events, numerous $Ca^{2+}$ signaling based molecular communications systems, models, and their characterization have been investigated and proposed over the years \cite{he2019calcium} \cite{allan2022encoding} \cite{barros2014transmission}.  

\subsection{Obtaining a Perceptron from $Ca^{2+}$ Signaling}
\begin{figure*}%
    \centering
    \begin{subfigure}{0.3\textwidth}
        \includegraphics[height=110px,  trim = {0 0cm 0 0cm}, clip]{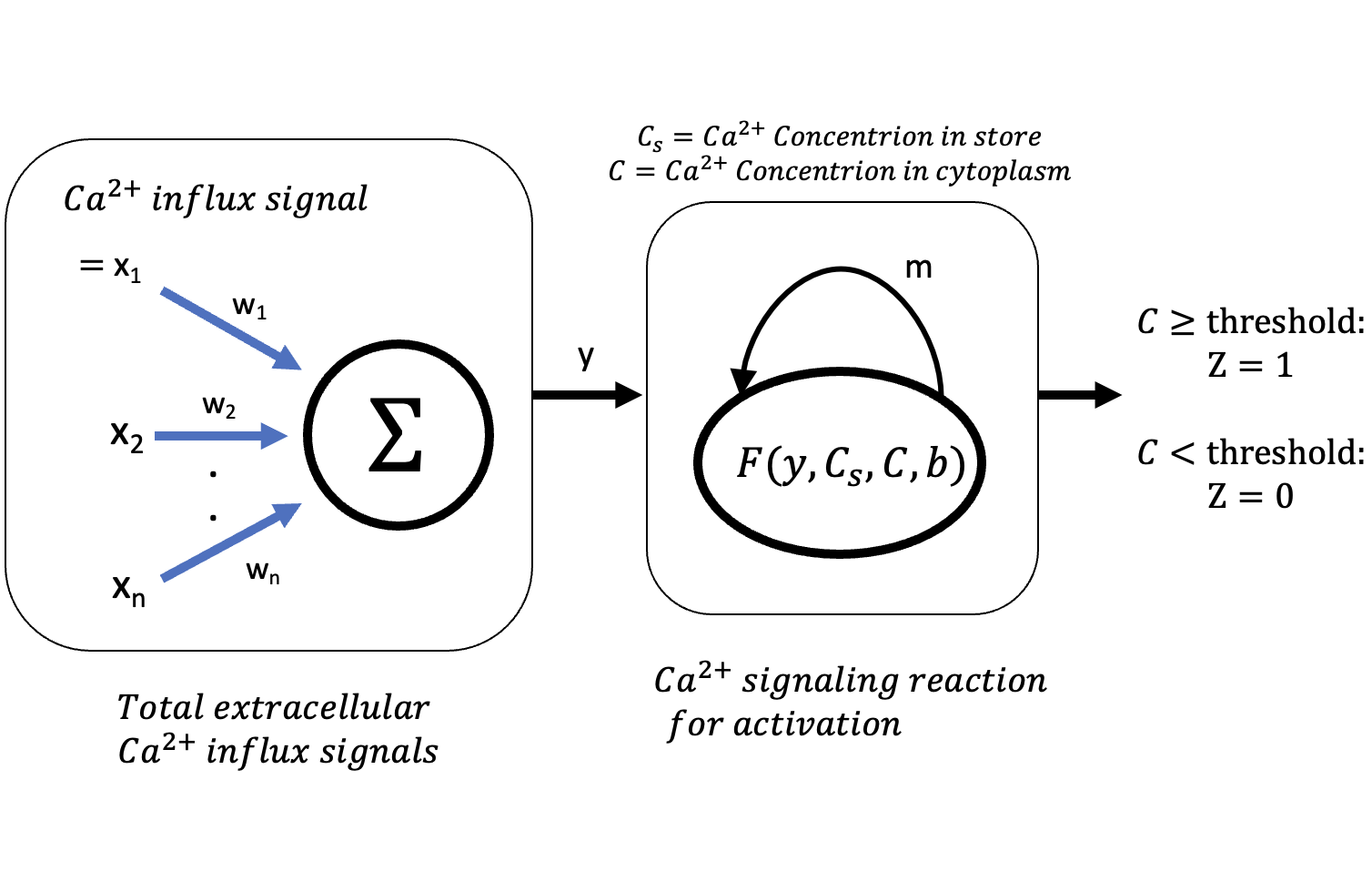}%
        \caption{}
        \label{fig:P_model}
    \end{subfigure}\hfill%
    \begin{subfigure}{0.2\textwidth}
        \includegraphics[height=120px,  trim = {0 0cm 0 0cm}, clip]{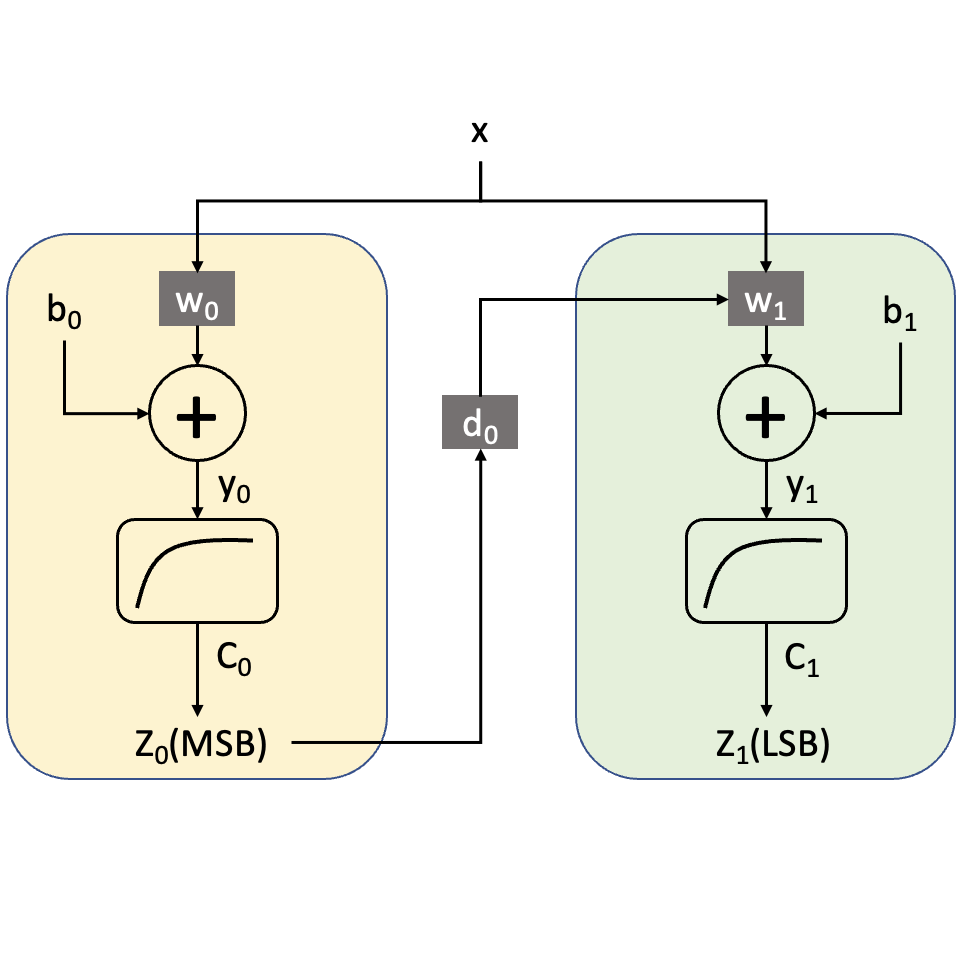}%
        \caption{}
        \label{fig:ADC_diagram}
    \end{subfigure}\hfill%
    \begin{subfigure}{0.4\textwidth}
        \includegraphics[height=110px, trim = {0cm 0cm 0 0cm}]{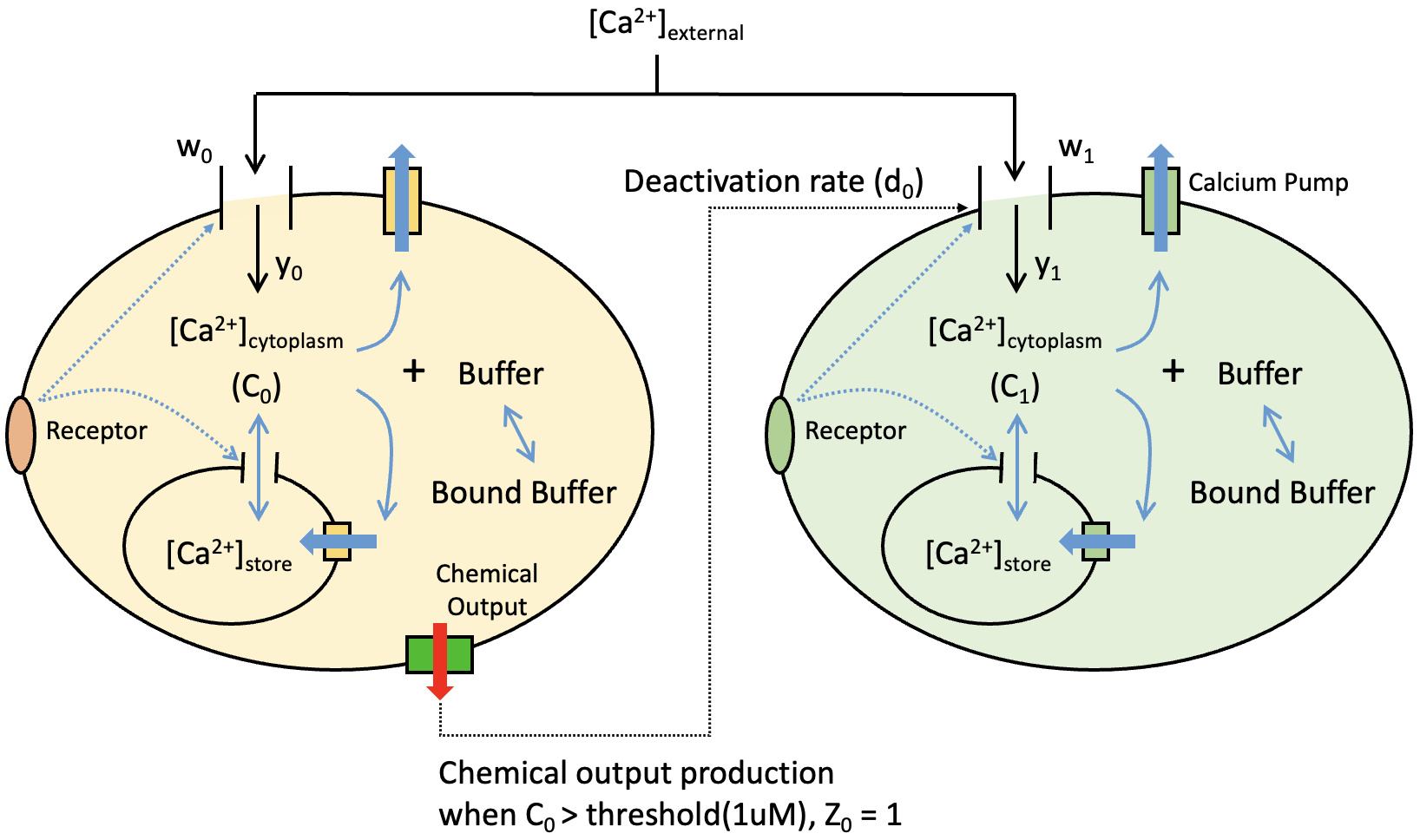}%
        \caption{}
        \label{fig:ADC_BioDiagram}
    \end{subfigure}%

    \begin{subfigure}{\textwidth}
        \includegraphics[width=\columnwidth, trim = {0cm 0cm 0 0cm}]{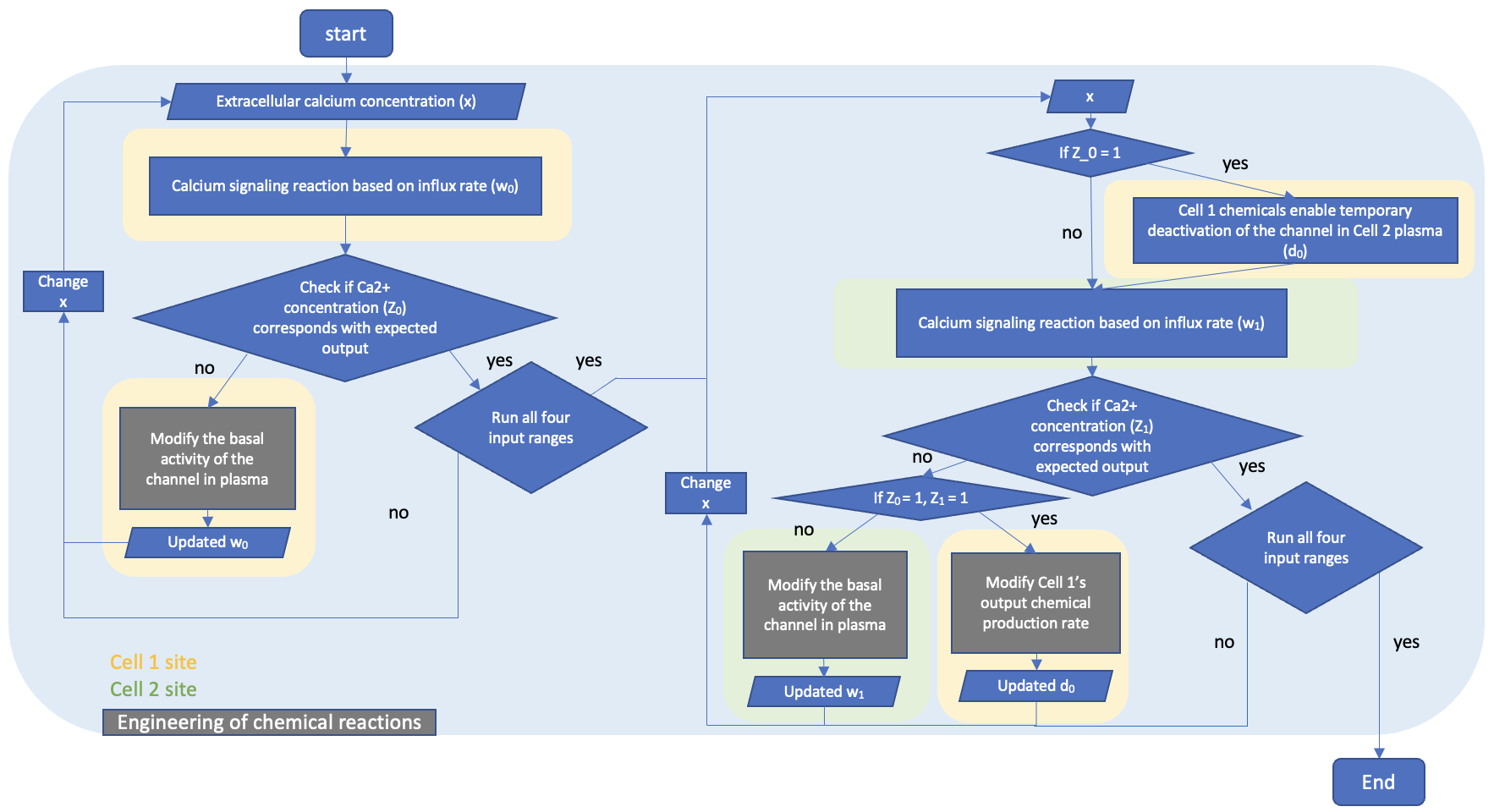}%
        \caption{}
        \label{fig:Ca_Flowchart}
    \end{subfigure}%

    \begin{subfigure}{.3\textwidth}
        \includegraphics[height=120px,  trim = {0 0cm 0 0cm}, clip]{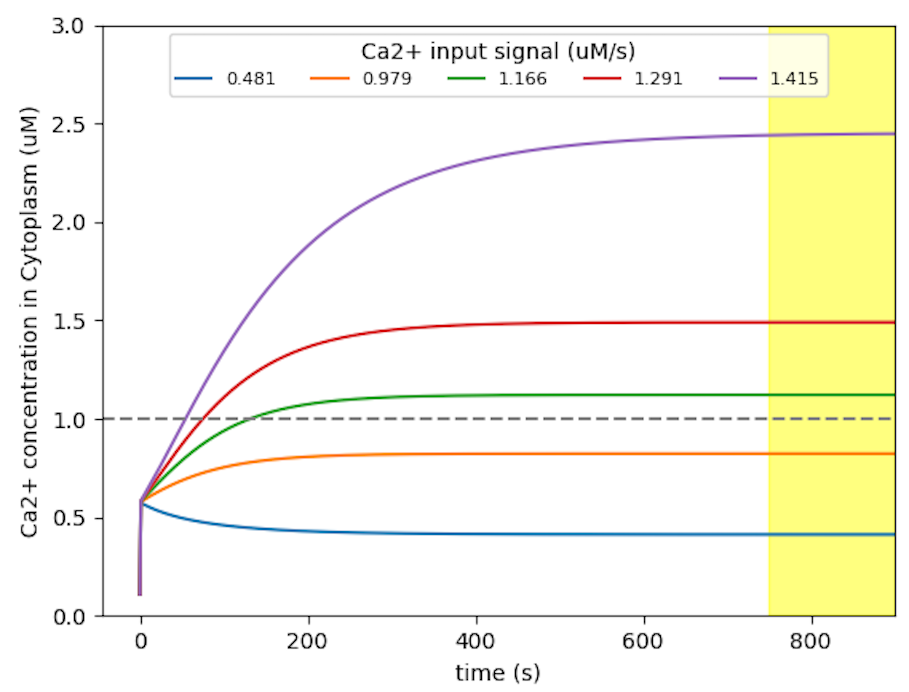}%
        \caption{}
        \label{fig:Ca_Model}
    \end{subfigure}\hfill%
    \begin{subfigure}{.28\textwidth}
        \includegraphics[height=140px,  trim = {0 0cm 0 0cm}, clip]{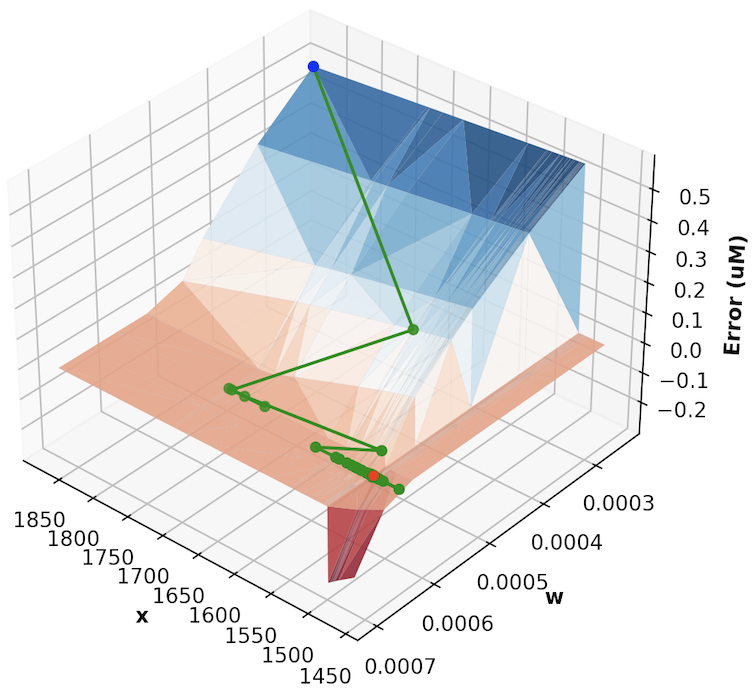}%
        \caption{}
        \label{fig:Cell1_train}
    \end{subfigure}\hfill%
    \begin{subfigure}{0.35\textwidth}
        \includegraphics[height=130px, trim = {0cm 0cm 0 0cm}]{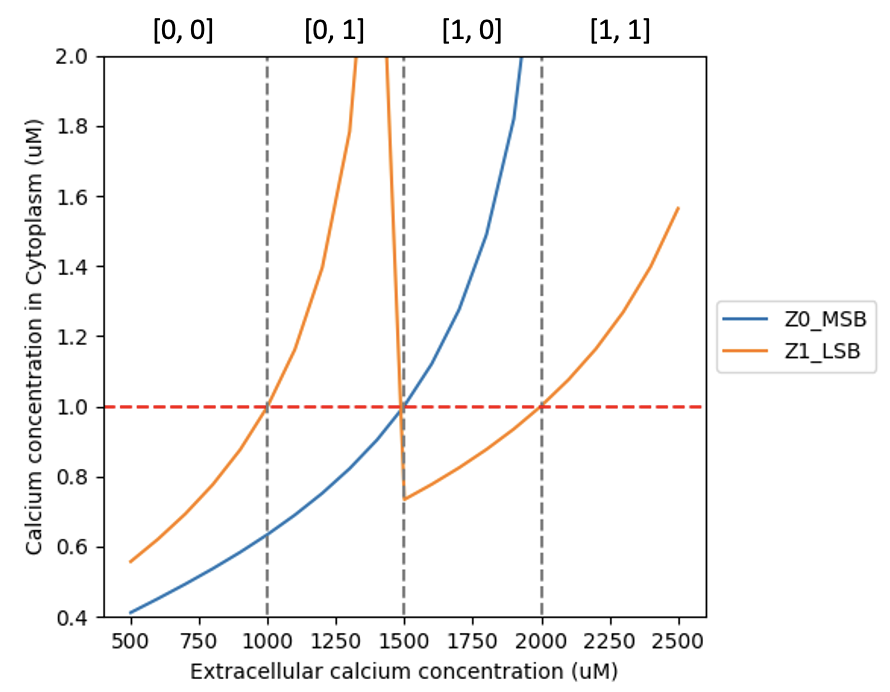}%
        \caption{}
        \label{fig:ADC_output}
    \end{subfigure}%
    \caption{Transforming $Ca^{2+}$ ions molecular communication into a perceptron. (a)A conventional perceptron model, (b)a Two-bit ADC architecture, (c) engineering $Ca^{2+}$ signaling into an ADC between two cells, (d) $Ca^{2+}$ signaling training process to modify the basal functional activity and communication channel flowchart, (e) Trained $Ca^{2+}$ ions transients in the cytoplasm, (f) Dynamics of Cell 1 weight $w_0$ through the training process with respect to the input extracellular $Ca^{2+}$ input ($x$), (g) variations in output $Ca^{2+}$ ions for the two cells to represent the ADC digital bits.}
    \label{fig:Calcium}
\end{figure*}

We adapt the Korngreen et al. model to exploit a $Ca^{2+}$ signaling system as a perceptron. As illustrated in Figure \ref{fig:P_model}, the input ($x$) will be the $Ca^{2+}$ ion concentration in the extracellular medium and the weight ($w$) is the $Ca^{2+}$ ions influx rate through the plasma membrane channels. Therefore, $x$*$w$ represents the amount of $Ca^{2+}$ ion influx ($y$) into the cytoplasm, representing its  transient. As described earlier, the $Ca^{2+}$ ion transients are multi-stage signaling processes that involve the transition of ions within the cytoplasm, store, buffer, as well as the extracellular medium and regulate the concentration in the cytoplasm. 
%The $F(y, C_s, C, b)$ is non-linear  $Ca^{2+}$ ions transient system function that regulates $Ca^{2+}$ ions inside the cell.
%The $Ca^{2+}$ ions transients results in an increasing linear function  %with an independent variable, $y$. As $y$ increases, $C$ increases. This in turn will allow us to use this $Ca^{2+}$ transients as a that allows us to exploit it as an  activation function. %The results of the $Ca^{2+}$ transients are shown in Fig. \ref{fig:Ca_Model} with various $Ca^{2+}$ influx signals($y$). \textcolor{blue}{Note: it's a non-linear function but has a positive relationship between input and output.  Repeating this function m times results in calcium signaling reaction for a certain time period in a cell, which regulates $Ca^{2+}$ ions inside the cell.  }
In order to train the $Ca^{2+}$ signaling process into a perception, the cell needs to be the incorporation of an engineered genetic circuit to modify its basal fractional activity to trigger the $Ca^{2+}$ signaling reaction or to modulate the influx channel. %channel on the plasma membrane by controlling the weight that represents the $Ca^{2+}$ influx rate.\textcolor{blue}{Note:weight is the default influx rate through channel with basal fractional activity. To control the weight, engineer needs to control the basal activity or do something on calcium channel.} 
In the case of a multiple-cell system to realize an ANN multi-perceptron network, the engineered genetic circuits are required to enable dynamic activation and deactivation of the $Ca^{2+}$ channel. %and this will be realized through an engineered circuit that invokes a chemical reaction. %{\bf you mean between cells?}.\textcolor{blue}{yes, in our case, Cell 1 chemicals temporarily deactivate the channel in cell 2 plasma.}

\subsection{Two-bit Analog to Digital Converter}
\subsubsection {Architecture} We adapted the $Ca^{2+}$ ion signaling model to create interacting perceptrons in multiple cells that altogether realize a two-bit  ADC through a simulation model. The architecture of a conventional ADC is illustrated in Figure  \ref{fig:ADC_diagram}. The equivalent model based on $Ca^{2+}$ signaling, where made clear the essential role of ion flow between two cells (the blue arrows in the Figure  \ref{fig:ADC_BioDiagram} indicate $Ca^{2+}$ ions reactions to facilitate this). The input $x$ is the incoming extracellular $Ca^{2+}$ concentration into the two cells, where the range of input considered in the simulation is set between $500\mu M$ to $2500\mu M$ and sampled according to an interval of $500\mu M$. By dividing this range into four intervals, each interval will produce different $Ca^{2+}$ signals from two cells, i.e., $Cell\ 1$ and $Cell\ 2$, which map to different digital bits. Based on this, the $Cell\ 1$ and $Cell\ 2$ produce the Most Significant Bit (MSB) and the least significant bit (LSB), respectively. %The ADC results are shown in Fig. \ref{fig:ADC_output}.%
$Ca^{2+}$ ions in the extracellular medium($x$) flow into the cytoplasm through the $Ca^{2+}$ channel with an influx rate $w_0$ and $w_1$ for  $Cell\ 1$ and $Cell\ 2$, respectively. A bias to the $Ca^{2+}$ ions influx for each of the two cells ($y_0$, $y_1$) is randomly selected and applied (in this example this is $b_0 = 0.169255\mu M$ and $b_1 = 0.287264\mu M$, respectively). Through the $Ca^{2+}$ transients, the ion concentrations in the cytoplasm that are set to $C_0$ and $C_1$, respectively. By setting a threshold, in our case, $1\mu M$, the $Ca^{2+}$ concentration in the cytoplasm can be converted into digital bit ($Z_0$, $Z_1$), which are the MSB and LSB. In order to make an ADC, $Cell\ 1$ is genetically engineered to produce molecules when enough $Ca^{2+}$ ions ($1\mu M$) are present in the cytoplasm. The output molecules temporally deactivate the calcium channel in $Cell\ 2$ plasma. This deactivation rate is indicated as $d_0$.

%In our case, the $Cell 1$ is artificially engineered for the ADC design to produce a chemical that deactivates a calcium channel on the $Cell 2$ plasma membrane only when there are enough $Ca^{2+}$ ions in the $Cell 1$’s cytoplasm (in other words when $Z_0$ is 1).The deactivation due to the $Cell 1$ output chemical has a rate of $d_0$. 
%As a membrane receptor in the model, the activation term $(1- d_0)$ is designed linearly.

\subsubsection{Training process}
The flow chart for training the $Ca^{2+}$ signaling perceptron is presented in Figure \ref{fig:Ca_Flowchart}. The two cells have to be trained to obtain optimal $Ca^{2+}$ influx rates ($w_0$, $w_1$) as well as the correct $Cell\ 1$'s calcium channel deactivation rate for $Cell\ 2$ ($d_0$) so that $Cell\ 1$ and $Cell\ 2$ can produce the aforementioned MSB and LSB, respectively. $Cell\ 1$ is trained first to find an optimal $w_0$, then $Cell\ 2$ to obtain $w_1$ and $d_0$. With initial $w_0$, $Ca^{2+}$ flows into $Cell\ 1$ and is regulated in the cytoplasm ($C_0$) for a certain period. Based on the amount of input from the extracellular medium ($x$), the concentration at saturation will represent an MSB digital bit ($Z_0$).
%According to the $Ca^{2+}$ transients model, $Cell 1$ regulates the calcium concentration in the cytoplasm. After a certain period, an engineer reads $C_0$ and converts it to $Z_0$.
When $Z_0$ is bit 0, but the expected output is bit 1, an activation chemical from the engineered circuit is injected to elevate the basal activity of the calcium channel in $Cell\ 1$ plasma. Due to the increased activity of the channel, an increased amount of $Ca^{2+}$ ions will flow into $Cell\ 1$, which means the influx rate ($w_0$) is also increased. 
%activate the $Cell 1$ calcium channel on the plasma membrane. Since the calcium channel is activated more,  is updated to a higher value.
For the opposite case, when $Z_0$ is bit 1 and the expected value is bit 0, a different deactivation  chemical signal is expressed by the engineered genetic circuit  to reduce the basal activity of the $Ca^{2+}$ channel. Then $w_0$ is updated to a lower value. Based on this sequential training process, the optimal $w_0$ will be found. %From the simulation, we found $w_0$ to be 0.0006230 $s^{-1}$. 
The same training process is  performed on $Cell\ 2$, except for one case. This exception case is when $Z_0$ and $Z_1$ are bits 1, but the expected $Z_1$ is bit 0, which will require manual intervention  to modify the rate of $Cell\ 1$ output chemical production instead of injecting chemicals. 
%$Cell 1$ deactivation chemicals allow a one-time deactivation, while injecting  artificial chemicals permanently changes the channel's activity. This modification will update the $d_0$. %Based on our simulations, the value of $w_2$ is found to be 0.0008156 $s^{-1}$ and that of $d_0$ is 0.4996.
Figure~\ref{fig:Ca_Model} shows how the perceptron behaves for  different levels of $Ca^{2+}$ within the cytoplasm based on varying extracellular influx. Figure~\ref{fig:Cell1_train} illustrates an example of convergence of weight $w_0$ during training with respect to the error for varying levels of extracellular input ($x$). Finally, Figure~\ref{fig:ADC_output} shows the variations of output from the two cells that represent digital bits from $Cell\ 1$ and $Cell\ 2$. For example, an input between $1000\mu M$ and $1500\mu M$ results in '01', where the 0 bit is from $Cell\ 1$ and 1 bit is from $Cell\ 2$.

\section{Challenges}
\label{future}
While we have identified solutions that enable non-neural cells to develop perceptron properties, or the exploitation of gene regulations to obtain ANN functionalities, there is still a number of challenges that needs to be addressed to move towards practical applications in the future, some important ones are discussed next.

\subsection{Controlling Molecular Communications in Molecular Machine Learning}
The MML that we have discussed so far are based on training and computing operations that stem from communications of molecules and chemical reactions. %This is for both GRAI as well as Cell-based AI. However, 
To develop MML systems processes matching the computational capabilities of silicon-based technologies, we will eventually need to consider multi-layer perceptron architectures. While the genetic engineering will possibly be the main enabling technology, specific challenges are as follows.
%this is possibly through engineering of multi-species bacterial communication that we have proposed as future directions, there are a number of challenges that needs to be considered. 
Firstly, since the training of the edge weights of molecular signals, which in our case is based on population control. Therefore, a mechanism is required to ensure that parallel changes in the bacteriome can be performed to modify the relative population of different species/strains in the system. This becomes more challenging when we consider $Ca^{2+}$ signaling between cells and in particular controlling the flow of ions through the gap junction of cells. 
%Secondly, we have started to witness the power of AI through multi-layer neural network  architectures that has high capability to learn and compute. 
%While this may be achievable using GRAI
Secondly, while GRAI might be inherently including multi-layer perceptrons, the question is how do we determine appropriate chemical inputs to express genes of the input nodes and at the same time detect expressions on specific output nodes. From a multi-bacterial species perspective, this will require engineering of cells with different receptors 
to detect diverse molecular signals from the previous layers. The cells will, therefore, need to have the ability to detect signals efficiently and operate in noisy environments. The other challenge is the ability to synchronize all transmissions as signals propagate between different layers. The latter challenge, can have an immense impact on the reliability of the resulting ANN. Since we have shown that multiple ANNs are  embedded in a GRN through a sub-network, the question is whether multiple parallel processing can be achieved through different gene expression paths.

\subsection{Bio-Hybrid AI}
The paradigm of the Internet of Bio-Nano Things \cite{akyildiz2015internet} includes the need to interconnect molecular communication systems to connect to the cyber-Internet by propagating information between the molecular and the electrical domains. This can be realized through an electro-chemical based Bio-cyber interfaces. While this can allow to detect chemical outputs from the MML, an issue arises when we want to 
%activate or reconfigure the MML
actively interact and reconfigure the MML system from the electrical domain. In particular, the challenge lies in the mechanism to reconfigure the weights. 

\subsection{Responsible AI in Molecular Machine Learning}
As AI continues to spread and weave into our everyday lives, besides developing sophisticated hardware and software, we are facing new and emerging ethical concerns has risen, which altogether call for the notion of responsible AI. Responsible AI aims to address the ethical and legal issues in regards to deployment as well as utilization of AI. This is already a major challenge in conventional AI, which is necessary to address to provide trust for the public in using the technology. This challenge will deepen further when AI is extended in living machines. This is particularly true, when we consider the potential applications of learning-based living machines for treating diseases, where they can potentially be deployed into the body or the environment. Another challenge is also the security aspect, in the similar manner that this is a challenge in conventional AI. 
  
\section{Conclusion}
\label{conc}

As our society embraces AI to play a part in our everyday lives, we are starting to witness various forms and algorithms that are embedded into devices with different computational capabilites. In this paper we investigate MML for Biological AI, where AI occurs in living systems and is based on information propagation through chemical reaction and molecule transport, i.e., molecular communications. We reviewed the current background in Biological AI. This is followed by our proposed directions of MML through the GRN, bacterial multi-species communication, as well as $Ca^{2+}$ signaling. We then discuss future possible directions for the molecular communications research.

% or


%\appendices

%\section{}

%\section*{Acknowledgment}


\ifCLASSOPTIONcaptionsoff
  \newpage
\fi



%\bibliographystyle{IEEEtran}
\bibliographystyle{ieeetr}
\bibliography{sample}


%\end{IEEEbiography}


%\newpage



%\vfill


\end{document}